\documentclass{PoS}

\usepackage{amsmath,amsfonts,latexsym,amssymb,hhline,stmaryrd,color,verbatim,graphicx,epstopdf,slashed,multirow}
\usepackage{slashed}
\usepackage{lineno}
\usepackage{color}
\usepackage[utf8]{inputenc}
\usepackage{booktabs}
\usepackage{subfig}
\usepackage{pifont}
\usepackage{url}

\newcommand\sss{\scriptscriptstyle}

\newcommand {\beq} {\begin{equation}}
\newcommand {\eeq} {\end{equation}}
\newcommand {\bea} {\begin{eqnarray}}
\newcommand {\eea} {\end{eqnarray}}

\definecolor{darkred}{rgb}{0.7, 0.0, 0.0}

\def\be{\begin{equation}}
\def\ee{\end{equation}}
\def\bsp#1\esp{\begin{split}#1\end{split}}

\newcommand{\OO}{\ensuremath{\mathcal{O}}}
\newcommand{\pdp}{\ensuremath{\phi^\dagger\phi}}
\renewcommand{\phi}{\ensuremath{\varphi}}

\newcommand{\bpm}{\begin{pmatrix}}      
\newcommand{\epm}{\end{pmatrix}}

\newcommand{\Op}[1]{\OO_{\sss #1}}
\newcommand{\Opp}[2]{\OO_{\sss #1}^{\sss #2}}

\newcommand{\red}[1]{ \textcolor{red}{#1} }
 \def\lra#1{\overset{\text{\scriptsize$\leftrightarrow$}}{#1}}

\title{Modified interactions in the top-quark electroweak sector: exploiting unitarity violating effects at the amplitude level to probe New Physics}

\ShortTitle{Modified interactions in the top-quark electroweak sector}

\author{\speaker{Luca Mantani}\\
        Center for Cosmology, Particle Physics and Phenomenology - CP3, Universite Catholique de Louvain, Louvain-la-neuve, Belgium\\
        E-mail: \email{luca.mantani@uclouvain.be}}

\author{Fabio Maltoni\\
        Center for Cosmology, Particle Physics and Phenomenology - CP3, Universite Catholique de Louvain, Louvain-la-neuve, Belgium\\
        E-mail: \email{fabio.maltoni@uclouvain.be}}

\author{Ken Mimasu\\
        Center for Cosmology, Particle Physics and Phenomenology - CP3, Universite Catholique de Louvain, Louvain-la-neuve, Belgium\\
        E-mail: \email{ken.mimasu@uclouvain.be}}

\abstract{We present a broad study of collider processes that embed $2 \to 2$ scattering amplitudes involving top quarks in the 
Electroweak sector. We parametrise the modified interactions using the Standard Model Effective Field Theory framework and discuss
how the New Physics effects lead to unitarity violating behaviour at the amplitude level. For each scattering amplitude we compute 
the helicity amplitudes in the high energy limit paying special attention to the effects of the higher dimensional operators.
We also discuss whether and to what extent the unitarity violating effects are retained in physical processes at colliders. 
}

\FullConference{
European Physical Society Conference on High Energy Physics - EPS-HEP2019 -\\
			10-17 July, 2019\\
			Ghent, Belgium}

\begin{document}

\section{Introduction}

The Standard Model (SM) is a spontaneously broken non-Abelian gauge Yukawa renormalisable theory.
As a consequence of that, the unitarity of the theory is granted by a set of intricate cancellations in the scattering amplitudes that would otherwise 
display unacceptable energy growths in the high energy limit. One the most famous examples of such cancellations is longitudinal
$W$ scattering~\cite{LlewellynSmith:1973yud,Lee:1977eg,Lee:1977yc} that exhibits an energy growing behaviour if the quartic gauge interaction and the Higgs contribution is not taken 
into account. This is just a consequence of the fact that the Higgs mechanism provides a consistent way of giving masses to the 
gauge bosons without explicitly breaking the gauge symmetry. Analogous reasoning can be applied to fermion scatterings
which display the same pattern of neat cancellations. In particular it can be shown~\cite{Appelquist:1987cf, Maltoni:2001dc, Maltoni:2000iq} that for processes involving fermions, 
the unitarity bounds scale with the inverse of the fermion mass and in this sense the top quark is the best probe we have to
look for deviations from the Standard Model .
Recently the observation of $t\bar{t}H$ production at the LHC has confirmed the Standard Model prediction that the top Yukawa is of order
one and opened the way for the precise determination of the top quark electroweak interactions. In fact as of today, the top quark couplings to the 
EW bosons are not precisely known. Their precise determination is one of the objective of the physics program of LHC and future colliders.
It seems therefore natural to understand how we can best constrain the top quark interactions with the goal of looking for
New Physics at colliders. In particular the observation of the aforementioned unitarity violating behaviours could be exploited
to assess the presence of heavy states beyond the reach of the collider, allowing us to infer indirectly the existence of new particles.
The Standard Model Effective Field Theory (SMEFT) is a suitable framework to parametrise these effects, describing deviations from the 
SM in a model independent manner. This framework is rooted in a gauge invariant description of new interactions through higher dimensional 
operators preserving the SM symmetries and it is mappable to a wide variety of Beyond Standard Model (BSM) theories. These operators lead to
both modified couplings and to new Lorentz structures and contact terms involving the SM fields. In this study~\cite{Maltoni:2019aot}, we interest ourselves 
in the high energy behaviour of a general class of EW scattering amplitudes involving a pair of fermions including at least one top quark
and two bosonic EW states, i.e., an EW gauge boson or the Higgs boson.
Other works have been performed in this direction. In particular in~\cite{Corbett:2014ora,Corbett:2017qgl} the derivation of
unitarity constraints on SMEFT operators in the case of massless limit is performed. In~\cite{Dror:2015nkp} the potential to exploit
non-standard top quark interactions is discussed in details and several of the collider processes that we study are also discussed.
In our work, detailed phenomenological analyses are not performed, with a broader, horizontal approach taken. Rather, we survey a considerable 
number of scatterings and associated collider processes focusing on quantifying the high energy behaviour, the associated sensitivity to 
SMEFT operators and discussing several general phenomenological and experimental issues.

\section{Theoretical framework}

\subsection{The Standard Model Effective Field Theory}

In order to parametrise the effects of New Physics, we focus on the SMEFT framework employing the Warsaw basis of operators~\cite{Buchmuller:1985jz,Grzadkowski:2010es} 
truncated at dimension six. Being only interested in the top quark electroweak sector, we impose a flavour symmetry $U(3)_{\ell}\times U(3)_{e}\times U(3)_{d}\times U(2)_{q}\times U(2)_{u}$, on our effective theory such that operators concerning deviations from top/third generation quark interactions 
can be singled out (see~\cite{AguilarSaavedra:2018nen} for a detailed classification). The labels $\ell,e,d,q,u$  refer to the fermionic representations of the SM: the lepton doublet, right handed lepton, right handed down-type quark, quark doublet and right handed up-type quark, respectively. The flavour symmetry considered only allows for a 
Yukawa interaction for the top quark meaning that all the other fermions are considered massless. In addition to that, the symmetry implies a flavour-universality for operators involving vector fermion currents. Only the operators that involve the 3rd generation quark doublet and right handed top quark can have independent coefficients.
Once we restrict ourselves to the EW sector of the top quark, we are left with the set of operators showed in Table~\ref{tab:operators}.

\begin{table}
{\centering
\renewcommand{\arraystretch}{1.4}
\begin{tabular}{|ll|ll|}
    \hline
     $\Op{W}$&
     $\varepsilon_{\sss IJK}\,W^{\sss I}_{\mu\nu}\,
                             {W^{{\sss J},}}^{\nu\rho}\,
                             {W^{{\sss K},}}^{\mu}_{\rho}$&
     $\Op{t\phi}$&
     $\left(\pdp-\tfrac{v^2}{2}\right)
     \bar{Q}\,t\,\tilde{\phi} + \text{h.c.}$
     \tabularnewline
     $\Op{\phi W}$&
     $\left(\pdp-\tfrac{v^2}{2}\right)W^{\mu\nu}_{\sss I}\,
                                    W_{\mu\nu}^{\sss I}$&
     $\Op{tW}$&
     $i\big(\bar{Q}\sigma^{\mu\nu}\,\tau_{\sss I}\,t\big)\,
     \tilde{\phi}\,W^I_{\mu\nu}
     + \text{h.c.}$
     \tabularnewline
     $\Op{\phi B}$&
     $\left(\pdp-\tfrac{v^2}{2}\right)B^{\mu\nu}\,
                                    B_{\mu\nu}$&
     $\Op{tB}$&
     $i\big(\bar{Q}\sigma^{\mu\nu}\,t\big)
     \,\tilde{\phi}\,B_{\mu\nu}
     + \text{h.c.}$
     \tabularnewline
     \cline{3-4}
     $\Op{\phi WB}$&
     $(\phi^\dagger \tau_{\sss I}\phi)\,B^{\mu\nu}W_{\mu\nu}^{\sss I}\,$&
     $\Op{\phi Q}^{\sss(3)}$&
     $i\big(\phi^\dagger\lra{D}_\mu\,\tau_{\sss I}\phi\big)
     \big(\bar{Q}\,\gamma^\mu\,\tau^{\sss I}Q\big)$
     \tabularnewline

     $\Op{\phi D}$&
     $(\phi^\dagger D^\mu\phi)^\dagger(\phi^\dagger D_\mu\phi)$&
     $\Op{\phi Q}^{\sss(1)}$&
     $i\big(\phi^\dagger\lra{D}_\mu\,\phi\big)
     \big(\bar{Q}\,\gamma^\mu\,Q\big)$
     \tabularnewline
     $\Op{\phi \square}$&
     $(\varphi^\dagger\varphi)\square(\varphi^\dagger\varphi)$ &
     $\Op{\phi t}$&
     $i\big(\phi^\dagger\lra{D}_\mu\,\phi\big)
     \big(\bar{t}\,\gamma^\mu\,t\big)$
      \tabularnewline
      &&     
      $\Op{\phi tb}$&
     $i\big(\tilde{\phi}^\dagger\,{D}_\mu\,\phi\big)
     \big(\bar{t}\,\gamma^\mu\,b\big)
     + \text{h.c.}$
     \tabularnewline
      \hline
      %

     %

 
  \end{tabular}

\caption{\label{tab:operators}
SMEFT operators describing new interactions involving the EW and top quark 
sectors, consistent with a $U(3)^3\times U(2)^2$ flavour symmetry.  
$Q,\,t$ and $b$ denote the third generation components of $q,\,u$ and $d$.
}
}
 \end{table}

In the operators definitions, the following conventions have been used:

\begin{align}
   \phi^\dag {\overleftrightarrow D}_\mu \phi&=\phi^\dag D^\mu\phi-(D_\mu\phi)^\dag\phi\\
   \phi^\dag \tau_{\sss K} {\overleftrightarrow D}^\mu \phi&=
   \phi^\dag \tau_{\sss K}D^\mu\phi-(D^\mu\phi)^\dag \tau_{\sss K}\phi \\
   W^{\sss K}_{\mu\nu} &= \partial_\mu W^{\sss K}_\nu 
   - \partial_\nu W^{\sss K}_\mu 
   + g \epsilon_{\sss IJ}{}^{\sss K} \ W^{\sss I}_\mu W^{\sss J}_\nu\\
   B_{\mu\nu} &= \partial_\mu B_\nu - \partial_\nu B_\mu \\
   D_\mu\phi =& \left(\partial_\mu -  i \frac{g}{2} \tau_{\sss K} W_\mu^{\sss K} - i\frac12 g^\prime B_\mu\right)\phi
 \end{align}
 where $\tau_I$ are the Pauli sigma matrices. Technically the operator $\Op{\phi tb}$ breaks the above mentioned flavour symmetry,
 but we decided to retain it in our study due to its unique right-handed charged current structure.
 The Universal FeynRules Output (UFO)~\cite{Degrande:2011ua} and the FeynArts~\cite{HAHN2001418} models used are available in the FeynRules models database~\cite{SMEFTatNLO}.

\subsection{Existing constraints on dimension-6 operators}

We report a table with up to date existing constraints on the Wilson coefficients that are of interest in our study in Table~\ref{tab:constraints}.
A clear hierarchy can be seen comparing the operators on the left column with the ones in the right column. The operators 
on the left do not explicitly involve a top quark field in their definitions and they are therefore best constrained from EW 
precision measurements, diboson and Higgs production processes. Those featuring a top quark on the other hand are constrained from
single top, top pair production and associated productions. For this reason we will mainly focus on the study of the latter, 
especially when looking at collider processes.

\setlength{\tabcolsep}{4pt}
\renewcommand{\arraystretch}{1.4}
\begin{table}[h!]
\begin{center} 
{\footnotesize
\begin{tabular}{|c|c|c|c|c|c|}
\hline
\multirow{2}{*}{Operator} & \multicolumn{2}{c|}{Limit on $c_i$ $\big[$TeV$^{-2}$$\big]$} & 
\multirow{2}{*}{Operator} & \multicolumn{2}{c|}{Limit on $c_i$ $\big[$TeV$^{-2}$$\big]$}   
\tabularnewline\cline{2-3}\cline{5-6}
{}                  & Individual    & Marginalised &   
{}                  & Individual    & Marginalised  
\tabularnewline\hline
 $\Op{\phi D}$            & [-0.021,0.0055]~\cite{Ellis:2018gqa}  & [-0.45,0.50]~\cite{Ellis:2018gqa}     &   
 $\Op{t \phi}$            & [-5.3,1.6]~\cite{Hartland:2019bjb}    & [-60,10]~\cite{Hartland:2019bjb}
\tabularnewline\hline
 $\Op{\phi \Box}$         & [-0.78,1.44]~\cite{Ellis:2018gqa}     & [-1.24,16.2]~\cite{Ellis:2018gqa}     &   
 $\Op{tB}$                & [-7.09,4.68]~\cite{Buckley:2015lku}   & $-$
\tabularnewline\hline
 $\Op{\phi B}$            & [-0.0033,0.0031]~\cite{Ellis:2018gqa} & [-0.13,0.21]~\cite{Ellis:2018gqa}     &   
 $\Op{tW}$                & [-0.4,0.2]~\cite{Hartland:2019bjb}   & [-1.8,0.9]~\cite{Hartland:2019bjb}
\tabularnewline\hline
 $\Op{\phi W}$            & [-0.0093,0.011]~\cite{Ellis:2018gqa}  & [-0.50,0.40]~\cite{Ellis:2018gqa}     &   
 $\Op{\phi Q}^{\sss (1)}$ & [-3.10,3.10]~\cite{Buckley:2015lku}   & $-$
\tabularnewline\hline
 $\Op{\phi WB}$           & [-0.0051,0.0020]~\cite{Ellis:2018gqa} & [-0.17,0.33]~\cite{Ellis:2018gqa}     &   
 $\Op{\phi Q}^{\sss (3)}$ & [-0.9,0.6]~\cite{Hartland:2019bjb}   & [-5.5,5.8]~\cite{Hartland:2019bjb}
\tabularnewline\hline
 $\Op{W}$                 & [-0.18,0.18]~\cite{Butter:2016cvz}    & $-$                                   &
 $\Op{\phi t}$            & [-6.4,7.3]~\cite{Hartland:2019bjb}   & [-13,18]~\cite{Hartland:2019bjb}
\tabularnewline\hline
 {}                       & {}                                    & {}                                    &  
 $\Op{\phi tb}$           & [-5.28,5.28]~\cite{Alioli:2017ces}    & [27,8.7]~\cite{Hartland:2019bjb}
\tabularnewline\hline
\end{tabular}
}

\end{center}
\caption{\label{tab:constraints}
Individual and marginalised 95\% confidence intervals on Wilson coefficients 
collected from a selection of global fits to Higgs, top and EW gauge boson data.}
\end{table}
\renewcommand{\arraystretch}{1.}

\section{High energy scatterings at colliders}

\subsection{$2 \to 2$ high energy scatterings}

Having set up the theoretical framework we move on to the analysis of the modified interactions in the top quark sector by
selecting generic $2 \to 2$ scattering amplitudes involving at least one top quark and EW bosons. We identified 10 scattering amplitudes of 
this kind, listed in Table~\ref{tab:amplitude_organisation} and for each of them we computed all of the helicity amplitudes in the
high energy limit defined by $s \sim -t >> v^2$, where $s$ and $t$ are the Mandelstam variables.

\begin{table}[h!]
\centering
\begin{tabular}{|r|p{4.5cm}|p{4.5cm}|}
\hline
&Single-top &Two-top ($t\bar{t}$) \tabularnewline\hline
w/o Higgs  &  $b \, W \to t \, (Z/\gamma)$ \hfill
& $t \, W \to t \, W$
\hfill \newline 
$t \, (Z/\gamma) \to t \, (Z/\gamma)$ \hfill  
\tabularnewline\hline
w/\phantom{o}  Higgs & $b \, W \to t \, h$ 
\hfill  & 
$t \, (Z/\gamma) \to t \, h$ \hfill \newline  
$t \, h \to t \, h$ \hfill \tabularnewline\hline
\end{tabular}
\caption{The ten $2\to 2$ scattering amplitudes whose high-energy behaviour we study in this paper.
\label{tab:amplitude_organisation}}
\end{table}

In particular we computed for each helicity configuration the leading energetic behaviour for the SM amplitude as well as for
the operators effects to the amplitude, collecting our results in various tables that can be consulted in~\cite{Maltoni:2019aot}.
As expected, all the SM amplitudes are at most constant in energy, while contributions from EFT operators have the maximum degree 
of growth $E^2$. It is interesting to observe that while most of the operators lead to a maximal energy growth in each scattering,
it is quite rare to have an operator that interferes in an energy growing way with the SM amplitude. This non-trivial behaviour is
particularly interesting because the interference term is linear in the Wilson coefficient and therefore is the leading 
effect in the EFT expansion. On the other hand, the quadratic contribution is guaranteed to grow with energy, but is more suppressed by
the scale of New Physics $\Lambda^4$. Singling out the scattering amplitudes and the operators that have this feature is helpful in order
to identify the collider processes that are most probably sensitive to modified interactions. A summary of our findings is
reported in Table~\ref{tab:summary_table}.

\begin{table}[]
\centering
\scriptsize
\begin{tabular}{|c|c|c|c|c|c|c|c|c|c|c|c|c|c|}
 \hline
                          & $\mathcal{O}_{\phi D}$ & $\mathcal{O}_{\phi d}$ &  $\mathcal{O}_{\phi BB}$ & $\mathcal{O}_{\phi W}$  & $\mathcal{O}_{\phi WB}$  & $\mathcal{O}_{W}$ & $\mathcal{O}_{t \phi}$ & $\mathcal{O}_{tB}$ & $\mathcal{O}_{tW}$ & $\mathcal{O}_{\phi Q}^{(1)}$ & $\mathcal{O}_{\phi Q}^{(3)}$ &  $\mathcal{O}_{\phi t}$ & $\mathcal{O}_{\phi tb}$ 
 \tabularnewline\hline
 
 $b\,W\to t\,Z$           & $E$           & $-$              & $-$            & $-$            & $E$             & $E^2$    & $-$           & $E^2$     & $E^2$     & $E$                      & \red{$E^2$}              & $E$            & $E^2$           
 \tabularnewline\hline

 $b\,W\to t\,\gamma$      & $-$           & $-$              & $-$            & $-$            & $E$             & $E^2$    & $-$           & $E^2$     & $E^2$     & $-$                      & $-$                      & $-$            & $-$           
 \tabularnewline\hline
 
 $b\,W\to t\,h$           & $-$           & $-$              & $-$            & $E$            & $-$             & $-$      & $E$           & $-$       & $E^2$     & $-$                      & \red{$E^2$}              & $-$            & $E^2$            
 \tabularnewline\hline
 
 $t\,W\to t\,W$           & $E$           & $E$              & $-$            & $E$            & $E$             & $E^2$    & $E$           & $E$       & $E^2$     & \red{$E^2$}              & \red{$E^2$}              & \red{$E^2$}    & $-$           
 \tabularnewline\hline
 
 $t\,Z\to t\,Z$           & $E$           & $E$              & $E$            & $E$            & $E$             & $-$      & $E$           & $E^2$     & $E^2$     & $E$                      & $E$                      & $E$            & $-$           
 \tabularnewline\hline
 
 $t\,Z\to t\,\gamma$      & $-$           & $-$              & $E$            & $E$            & $E$             & $-$      & $-$           & $E^2$     & $E^2$     & $-$                      & $E$                      & $-$            & $-$           
 \tabularnewline\hline
 
 $t\,\gamma\to t\,\gamma$ & $-$           & $-$              & $E$            & $E$            & $E$             & $-$      & $-$           & $E$       & $E$       & $-$                      & $-$                      & $-$            & $-$           
 \tabularnewline\hline
 
 $t\,Z\to t\,h$           & $E$           & $-$              & $E$            & $E$            & $E$             & $-$      & $E$           & $E^2$     & $E^2$     & \red{$E^2$}              & \red{$E^2$}              & \red{$E^2$}    & $-$           
 \tabularnewline\hline
  
 $t\,\gamma\to t\,h$      & $-$           & $-$              & $E$            & $E$            & $E$             & $-$      & $-$           & $E^2$     & $E^2$     & $-$                      & $-$                      & $-$            & $-$
 \tabularnewline\hline

$t\,h\to t\,h$            & $E$           & $E$              & $-$            & $-$            & $-$             & $-$      & $E$           & $-$       & $-$       & $-$                      & $-$                      & $-$            & $-$           
 \tabularnewline\hline
 
  \end{tabular}
\caption{Maximal energy growths induced by each operator on the set of scattering amplitudes considered. ``$-$'' denotes either no contribution or no energy growth and the red entries denote the fact that the interference between the SMEFT and the SM amplitudes also grows with energy.}
\label{tab:summary_table}
\end{table}

\subsection{Embedding the scattering amplitudes in collider processes}

\begin{figure}[h!]
\centering
\includegraphics[width=0.4\textwidth]{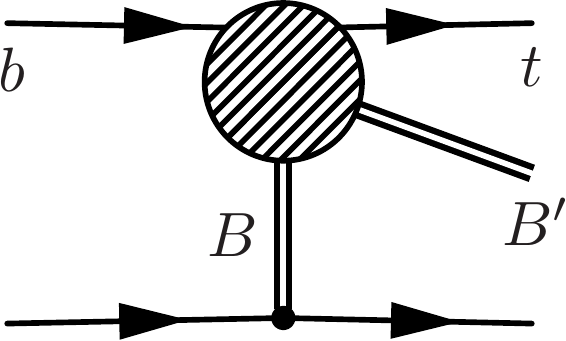}
\caption{\label{fig:topology}
Schematic Feynman diagram for the embedding of an EW top scattering amplitude into a physical, single-top process at a hadron collider. Here $f$ and $f^\prime$ must be a $b$- and $t$-quark respectively, while $B$ and $B^\prime$ can be several combinations of $Z,\,\gamma,\,W$ and $h$.
}
\end{figure}

In order to probe the top EW sector in physical processes and take advantage of the aforementioned unitarity violating effects,
the idea is to embed the $2 \to 2$ scattering amplitudes in real processes as pictured in Fig.~\ref{fig:topology} for the case of
single top production. With this in mind we turn to addressing the problem of what new information/sensitivity to the Wilson 
coefficients can be gained from collider measurements of such processes. In particular we investigate the degree to which the behaviour of the 
$2 \to 2$ scattering amplitude is preserved in going to higher multiplicities.
In our study we consider both present and possible future colliders, computing cross section computations with {\sc MadGraph5\_aMC@NLO}~\cite{Alwall:2011uj,Alwall:2014hca} using the SMEFT {\sc UFO} model based on the aforementioned {\sc FeynRules} implementation. We compute the linear and quadratic contributions to the
cross sections of each operator separately and define a naive sensitivity measure $r_i$ and $r_{i,i}$ by dividing by the pure EW SM
contribution. This relative impact has been computed by setting the Wilson coefficients to 1 TeV$^{-1}$. In order to gain insight on 
the high energy behaviour of the processes, we compute both the inclusive cross sections and the cross sections in a restricted high energy
region of phase space determined process by process by cuts on kinematical variables. The corresponding relative impacts are 
denoted by $r^\text{tot}$ and $r^\text{HE}$, respectively. A noticeable growth in the relative impact of the operator from inclusive to high-energy phase space can be a sign 
of unitarity violating behaviour due to the operator insertions.

\subsection{An interesting process: $tZW$}

\begin{figure}[h!]
  \centering \includegraphics[width=\linewidth]{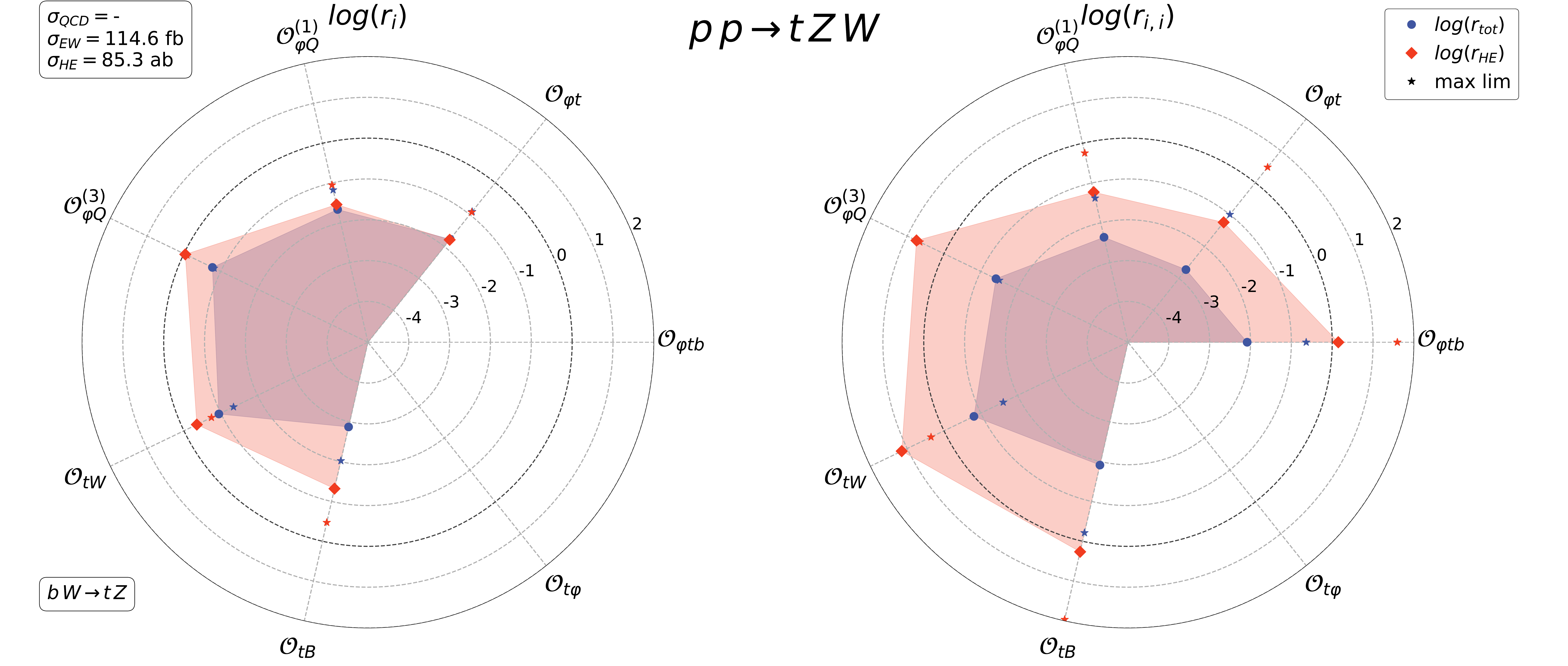}\\[1.5ex]
  \caption{Radar plot for the process $p \, p \to t \, Z \, W$ at the 13 TeV LHC.
 \label{fig:radar_tzw_LHC13}}
\end{figure}

Among the many processes studied in~\cite{Maltoni:2019aot}, we found $tZW$ to be a particularly sensitive process to modified 
interactions induced by effective operators on $b W \to t Z$ scattering. In Fig.~\ref{fig:radar_tzw_LHC13}, we report in a
compact format our results. In the top left corner we show the inclusive EW SM cross sections, while in the bottom left we recall which 
sub-amplitude is probed. We plot the ratios $r_i$ and $r_{i,i}$ in logarithmic scale for each operator. In the case of the linear term, 
being non-positive definite, we plot the absolute value. The blue dots represent the value of the ratios at inclusive level, 
while the red in the high energy region of phase space, which is defined by requiring both the $Z$ and the $W$ to have a 
$p_T > 500$ GeV. Finally the stars represent the absolute size of the impact once the Wilson coefficients limits in Table~\ref{tab:constraints}
are saturated.

As it can be seen from the radar plot, the process is quite sensitive to the operators at inclusive level for both linear and quadratic contributions.
Most importantly it exhibits energy growth at interference level for $\Opp{\phi Q}{(3)}$, which was expected by looking at Table~\ref{tab:summary_table}.
We find this process to be very promising, especially from a theoretical perspective. From an experimental point of view, the process cross section is
about five times smaller than $tZj$. In addition to that there are challenges in distinguishing it from $t\bar{t}Z$, but we think it would be very 
valuable to measure it in the context of globally constraining the SMEFT in the top sector.

\section{Conclusions}

We have presented a comprehensive study of the energy growing effects in the top quark EW sector by parametrising the BSM
effects in the SMEFT framework. We did so by assessing the impact of the higher dimensional operators on the helicity amplitudes 
for a set of $2 \to 2$ scattering amplitudes involving the top quark. We found that almost all of them lead to maximal energy growth
$E^2$ and we identified the cases in which these contributions constructively interfere in an energy growing way with the SM counterpart.
These contributions in particular are very promising to investigate at colliders, for which we have performed a detailed study,
reporting here the interesting case of $tZW$, which is a potentially good probe of the $b W \to t Z$ scattering. It is worth pointing
out that each interesting process will merit a dedicated phenomenological study to assess the true sensitivity to the Wilson coefficients.
These will have to take into account the presence of QCD background as well as process dependent reconstruction efficiencies in addition
to the parametric sensitivity that we have explored in this work.

\section*{Acknowledgements}
This work has received funding from the European Union's Horizon 2020 research and innovation programme as part of the Marie Skłodowska-Curie Innovative Training Network MCnetITN3 (grant agreement no. 722104).


\bibliographystyle{utphys}
\bibliography{refs.bib}
\end{document}